\documentclass[a4paper,referee,fleqn]{mnras} 

\usepackage[T1]{fontenc}
\usepackage{ae,aecompl}

\usepackage{graphicx}	
\usepackage{amsmath}	
\usepackage{amssymb}	

\begin{document}
\title[Basaltic material beyond Vesta]{Basaltic material in the main belt: a tale of two (or more) parent bodies?}


\author[S. Ieva et al.]{
S. Ieva,$^{1}$\thanks{E-mail:simone.ieva@oa-roma.inaf.it}
E. Dotto,$^{1}$
D. Lazzaro,$^{2}$
D. Fulvio,$^{3}$
D. Perna,$^{1,4}$
E. Mazzotta Epifani,$^{1}$ \newauthor
H. Medeiros,$^{2}$
and M. Fulchignoni$^{4}$
\\
$^{1}$INAF -  Osservatorio Astronomico di Roma, via Frascati 33, I-00078 Monteporzio Catone (Roma), Italy\\
$^{2}$Observatorio Nacional, R. Gen. Jos\'e Cristino, 77 - S\~ao Crist\'ov\~ao, Rio de Janeiro - RJ, 20921-400, Brazil\\
$^{3}$Departamento de Fis\'ica, Pontif\'icia Universidade Cat\'olica do Rio de Janeiro, Rua Marques de S\~ao Vicente 225, 22451-900 Rio de\\ Janeiro, Brazil\\
$^{4}$LESIA, Observatoire de Paris, PSL Research University, CNRS, Univ. Paris Diderot, Sorbonne Paris Cit\'e, UPMC Univ., Paris 06, \\ Sorbonne Universit\'e, 5 Place J. Janssen, Meudon Cedex F-92195, France\\
}

\date{Accepted 2018 June 11. Received 2018 June 7; in original form 2018 April 6}

\pubyear{2018}

\label{firstpage}
\pagerange{\pageref{firstpage}--\pageref{lastpage}}
\maketitle

\begin{abstract}
The majority of basaltic objects in the main belt are dynamically connected to Vesta, the largest differentiated asteroid known. Others, due to their current orbital parameters, cannot be easily dynamically linked to Vesta. This is particularly true for all the basaltic asteroids located beyond 2.5 au, where lies the 3:1 mean motion resonance with Jupiter.

In order to investigate the presence of other V-type asteroids in the middle and outer main belt (MOVs) we started an observational campaign to spectroscopically characterize in the visible range MOV candidates. We observed  18 basaltic candidates from TNG and ESO - NTT between 2015 and 2016.  We derived spectral parameters using the same approach adopted in our recent statistical analysis and we compared our data with orbital parameters to look for possible clusters of MOVs in the main belt, symptomatic for a new basaltic family. 

Our analysis seemed to point out that MOVs show different spectral parameters respect to other basaltic bodies in the main belt, which could account for a diverse mineralogy than Vesta; moreover, some of them belong to the Eos family, suggesting the possibility of another basaltic progenitor.
This could have strong repercussions on the temperature gradient present in the early Solar System, and on our current understanding of differentiation processes.
\end{abstract}




\section{Introduction}


The study of basaltic asteroids in the main belt has been a powerful tool to constrain the presence and frequency of differentiated material in the early Solar System. These asteroids, classified as V-type in all the latest taxonomies (Tholen \& Barucci 1989, Bus \& Binzel 2002, DeMeo et al. 2009) are thought to represent the crust of planetesimals that undergone a complete metal-silicate differentiation: iron core, olivine mantle and basaltic crust.
The spectrum of a basaltic asteroid can be easily identified by the presence of two deep absorption bands, near 0.9  and 1.9 $\mu m$. The same spectrum has been also identified in a type of  achondrite meteorites: the Howardites, Eucrites and Diogenites (collectively known as HED meteorites) for which Vesta, the first basaltic asteroid identified by McCord et al. (1970),  has been considered the parent body.

The identification of several basaltic V-type asteroids in orbits close to Vesta (Binzel \& Xu, 1993) and the discovery of a giant crater in the south pole of Vesta (Thomas et al. 1997) completed the picture.
The classical scenario foresees that one or more impacts on the surface of Vesta created a  swarm of basaltic fragments (i.e. the\emph{ vestoids}), forming the Vesta dynamical family. Some of them, injected into strong resonances, became Near-Earth Asteroids (NEAs, Cruikshank et al. 1991) and finally collide with our planet, being recovered as  HED meteorites.

The discovery of V-type asteroids with no dynamical link with Vesta, beyond the 3:1 mean motion resonance with Jupiter (Lazzaro et al. 2000, Binzel et al. 2006, Duffard \& Roig 2009) raised doubts if all the basaltic asteroids in the Solar System come from Vesta.
Dynamical simulations show that the probability for an asteroid of a D$>$ 5 km to evolve from the Vesta family and cross over the 3:1 resonance, reaching a stable orbit in the middle belt, is almost 1\% (Roig et al. 2008). Moreover,  laboratory studies on meteorites (Bland et al. 2009, Scott et al. 2009) and dynamical considerations (Carruba et al. 2014) suggest that several large asteroids (D $\sim$ 150-300 km)  should have been differentiated in the early Solar System. The inconclusive search of these bodies lead to the idea that these basaltic progenitors were battered to bits; 
or that maybe our understanding of differentiation processes is not complete (Lazzaro 2009).

At the moment, only 5 Middle and Outer belt V-types (MOVs) have been characterized with both visible and near-infrared spectroscopic observations.
Our recent statistical analysis on the largest sample of V-type spectra available in literature (Ieva et al. 2016) has pointed out that MOVs have spectral parameters that seem to differ from basaltic objects linked to Vesta. Furthermore,  spectral parameters for at least two MOVs (Magnya and Panarea) doesn't fit in the complete spectral characterization of Vesta retrieved by the DAWN mission (Ammannito et al. 2013).
Finally, these two MOVs present a size that seem incompatible with an excavation from the basin found on the south pole of Vesta (Marchi et al. 2012). 

In order to characterize via visible spectroscopy  basaltic candidates in the middle/outer main belt we conducted two observational campaigns at  Telescopio Nazionale Galileo (TNG) in 2015 and at ESO - New Technology Telescope (NTT) in 2016. 
The observed objects were identified among the SDSS - Moving Object Catalog (MOC), making the assumption that candidates with photometric colors and albedo indicative of a basaltic composition are indeed basaltic asteroids. They were selected among different databases of putative V-type asteroids (Roig \& Gil-Hutton 2006, Masi et al. 2008, Carvano et al. 2010, Oszkiewicz et al. 2014). 

The paper is organized as follows: in Section 2 we report the instrumental setup and the reduction process; in Section 3 we show our data analysis, obtained using spectral parameters and orbital elements; finally, in Section 4 we interpret our results in the light of the recent advancements in the field and we expose our conclusions.

\section{Observations and data reduction}

At TNG data were obtained using the DOLORES instrument, equipped with the LR-R grism (0.45 - 1.00 $\mu$m range) and the 2 arcsec slit, oriented along the parallactic angle to avoid atmospheric differential refraction. At ESO - NTT, spectra were acquired using the EFOSC2 instrument, equipped with the grism \#1 (covering the 0.32 - 1.09 $\mu$m range) and the 2 arcsec slit, also oriented along the parallacting angle.  

Data reduction was performed with ESO-Midas software using standard procedures (see e.g. Ieva et al. 2014): 
\begin{itemize}
\item raw flux exposures were bias-subtracted and flat-field corrected;
\item spectra were integrated along the spatial axis between 1.5 and 2 FWHM, and corrected for the sky contribution;
\item spectra were calibrated in wavelength using several spectral lines obtained from calibration lamps;
\item spectra were corrected for extinction of the observing site and normalized at 0.55  $\mu$m;
\item finally, the reflectivity for the asteroids was obtained by dividing the spectrum of each target by the spectrum of one solar analogue (usually the one with the closest airmass of the object).
\end{itemize}

\begin{table*}
       \caption{Observational conditions  for 18 MOVs considered in this work}
        \label{observations}
\begin{tabular}{|l|c|c|c|c|c|c|c|} \hline
\hline
&\\
Object&	Date&		$\alpha$&	Telescope/Instrument&	$T_{exp} (s)$&	Airmass&	Solar analogue (Airm.) \\
\hline
10769&	14/04/2015&		4.6&	TNG/DOLORES&	3x300&	1.32&	 SA107-998 (1.34)\\
14447&	09/04/2016&		7.8&	ESO-NTT/EFOSC2&	1200&		1.18&	 SA107-684 (1.18)\\
14562&	09/04/2016&		2.2&	ESO-NTT/EFOSC2&	300&		1.31&	 SA98-978 (1.24)\\
22308&	09/04/2016&		8.1&	ESO-NTT/EFOSC2&	600&		1.22&	 SA98-978 (1.24) \\
23321&	14/04/2015&		20.0&	TNG/DOLORES&	3x600&	1.54&	 SA102-1081 (1.52)\\
24264&	14/04/2015&		5.9&	TNG/DOLORES&	3x500&	1.37&	  SA107-998 (1.34)\\
27219&	28/04/2015&		19.9&	TNG/DOLORES&	3x300&	1.18&	 SA102-1081 (1.18)\\
41243&	09/04/2016&		17.2&	ESO-NTT/EFOSC2&	1200&		1.29&	 SA98-978 (1.24)\\
44496&	09/04/2016&		13.6&	ESO-NTT/EFOSC2&	180&		1.44&	 SA107-998 (1.51)\\
46245&	09/04/2016&		6.0&	ESO-NTT/EFOSC2&	600&		1.15&	 SA107-684 (1.18)\\
47063&	14/04/2015&		7.0&	TNG/DOLORES&	3x600&	1.48&	 BS4486 (1.53)\\
48448&	09/04/2016&		9.3&	ESO-NTT/EFOSC2&	300&		1.26&	  SA98-978 (1.24)\\
52002&	14/04/2015&		7.9&	TNG/DOLORES&	3x500&	1.49&	  BS4486 (1.53)\\
63256&	14/04/2015&		20.9&	TNG/DOLORES&	1200&		1.50&	  SA102-1081 (1.52)\\
77695&	09/04/2016&		4.8&	ESO-NTT/EFOSC2&	600&		1.11&	  SA102-1081 (1.15)\\
81854&	09/04/2016&		5.4&	ESO-NTT/EFOSC2&	900&		1.45&	 SA107-998 (1.51)\\
87128&	09/04/2016&		8.3&	ESO-NTT/EFOSC2&	900&		1.20&	 SA98-978 (1.24)\\
114544&	09/04/2016&		11.0&	ESO-NTT/EFOSC2&	300&		1.27&	 SA98-978 (1.24)\\
\hline													
\hline	
\end{tabular}

\textbf{NOTE:} $\alpha$ is the phase angle.

~\\
\raggedright
\smallskip
\end{table*}

We spectroscopically characterized 18 MOV candidates in the visible range (See  Tab. 1 for the observational conditions).
Objects were classified as V-type objects using the M4AST tool\footnote{http://m4ast.imcce.fr/index.php/index/start}, which uses a standard curve-matching technique to compare the input spectrum to all the taxonomic classes considered in the latest taxonomy (DeMeo et al. 2009). In Fig. 1 we report the obtained spectra, along with the normalized reflectance derived from the asteroid g r i z photometry, taken from the SDSS-MOC catalogue. As can be noted, some of the spectra show a great fringing beyond 0.8 $\mu$m and their classification as V-type must be taken with care, since it relies only on the linear part and not on the presence and position of the 0.9 $\mu$m absorption band.

 \begin{figure}
   \includegraphics[angle=0,width=15cm]{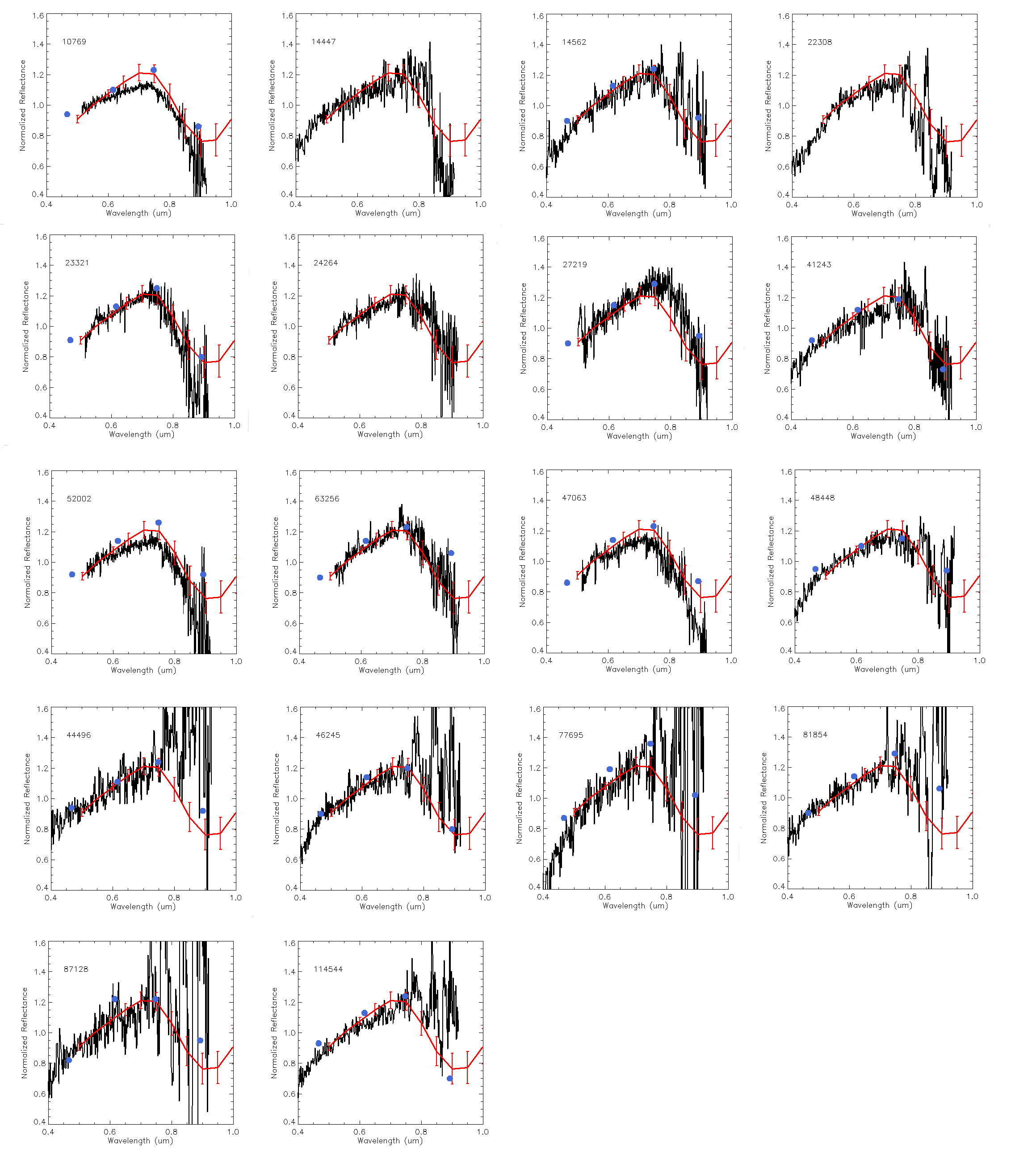}
      \caption{Visible spectra for our sample of MOVs characterized at TNG and ESO-NTT. The data has been median filtered and normalized at 0.55  $\mu$m, with the DeMeo et al. (2009) V-type taxonomy template overplotted in red along with the normalized g r i z reflectance (blue dots) at 0.467, 0.616, 0.748 and 0.893 $\mu$m, respectively. Spectra for 6 objects (44496, 46245, 77695, 81854, 87128, 114544) suffer from a great fringing beyond 0.8  $\mu$m. For these objects we were able to characterize only their reflectivity gradient between 0.5-0.75  $\mu$m (\emph{slopeA} in the text).}
         \label{spettri}
   \end{figure}

\section{Data analysis}
\subsection{Spectral parameters}

In order to characterize the spectral behavior of our sample of 18 MOVs and compare the results with our previous statistical analysis (Ieva et al. 2016) we decided to use the same approach and computed three spectral parameters: reflectivity gradients between 0.5-0.75 $\mu$m  (\emph{slopeA}) and 0.8-0.92 $\mu$m  (\emph{slopeB}) and the reflectance ratio 0.75/0.9 $\mu$m (\emph{apparent depth}, hereafter simply \emph{depth}).  For more details see Lazzaro et al. (1999).

The obtained spectral parameters are reported in Tab.2. As mentioned above, some spectra show great fringing afterwards 0.8 $\mu$m, making rather challenging the retrieval of \emph{slopeB} and \emph{depth}. For these asteroids (44496, 46245, 77695, 81854, 87128 and 114544) we were able to derive with enough confidence only the \emph{slopeA} parameter; they are therefore excluded from the subsequent analysis.

\begin{table*}
       \caption{Spectral and orbital parameters for our sample of MOVs characterized in this work. For completeness, we also reported MOVs previously analyzed in Ieva et al. (2016), together with average values for MOVs belonging to the three dynamical regions defined in Sect. 3.2, the whole sample of MOVs, Vesta family objects and S-type bodies computed from Lazzaro et al. (2004). }
        \label{parameters}
\begin{tabular}{|l|r|c|c|c|c|c|c|} \hline
\hline
Object&	SlopeA  &	Depth&	SlopeB &	 SlopeAcorr &	\emph{a} (au)&	\emph{e}&	\emph{i} (deg) \\
&	 ($\%/10^3 \textup{\AA})$ &	&	 ($\%/10^3 \textup{\AA})$&	 ($\%/10^3 \textup{\AA})$&	&	&	 \\
\hline
MOVs in Ieva et al. (2016)	&				&				&				&	&		&		&	\\
1459$^1$	&	15.55	$ \pm$ 0.16	&	1.81	$ \pm$ 0.06	&	-32.48 $ \pm$ 0.60	&	11.92	&	3.15&		0.21	&	15.37\\
10537	 &	8.79	$ \pm$ 0.39	&	1.50	$ \pm$ 0.01	&	-28.23 $ \pm$ 0.58	&	12.69	&	2.85&		0.10	&	6.32\\
21238	 &	10.09	$ \pm$ 0.15	&	1.80	$ \pm$ 0.25	&	-23.83 $ \pm$ 1.30	&	13.09	&	2.54&		0.13	&	10.76\\
40521$^2$	 &	9.99	$ \pm$ 0.19	&	1.78	$ \pm$ 0.29	&	-26.55 $ \pm$ 1.66	&	10.59	&	2.53&		0.05	&	12.46\\
\hline	
MOVs featured in this work	&				&				&				&	&		&		&	\\
10769$^3$&		6.42	$ \pm$ 0.26	&	1.82	$ \pm$ 0.14	&	-45.24 $ \pm$ 1.73	&	10.67	&	3.07	&	0.07	&	9.27	\\
14447$^3$	&	9.90	$ \pm$ 0.88	&	2.22	$ \pm$ 0.29	&	-29.15 $ \pm$ 6.01&	11.14	&	3.01	&	0.26	&	8.62	\\
14562$^1$	&	12.04	$ \pm$ 0.60	&	1.49	$ \pm$ 0.31	&	-30.71 $ \pm$ 6.83	&	10.31	&	3.12	&	0.18	&	16.34	\\
22308$^2$	&	9.95	$ \pm$ 0.50	&	2.00	$ \pm$ 0.24	&	-29.14 $ \pm$ 7.05	&	11.19	&	2.77	&	0.11	&	12.23	\\
23321$^2$	&	11.08	$ \pm$ 0.30	&	1.88	$ \pm$ 0.19	&	-47.88 $ \pm$ 2.84	&	12.96	&	2.77	&	0.12	&	13.58	\\
24264	&		9.03	$ \pm$ 0.44	&	1.74	$ \pm$ 0.15	&	-31.29 $ \pm$ 3.03	&	10.86	&	2.77	&	0.14	&	9.00	\\
27219$^1$	&	16.41	$ \pm$ 0.48	&	1.83	$ \pm$ 0.22	&	-55.15 $ \pm$ 2.68	&	12.95	&	3.12	&	0.19	&	16.75	\\
41243$^3$	&	8.98	$ \pm$ 0.65	&	1.82	$ \pm$ 0.31	&	-32.07 $ \pm$ 5.32&	12.54	&	2.99	&	0.08	&	10.95	\\
44496	&		11.67	$ \pm$ 0.87	&		-	&			-	&		12.01	&	3.09	&	0.09	&	13.93	\\
46245	&		10.32	$ \pm$ 0.62	&		-	&			-		&	10.87	&	3.09	&	0.08	&	10.42	\\
47063	&		9.19	$ \pm$ 0.66	&	1.82	$ \pm$ 0.13	&	-48.00 $ \pm$ 1.95	&	11.02	&	2.91	&	0.06	&	2.14	\\
48448	&		10.03	$ \pm$ 0.55	&	1.58	$ \pm$ 0.10	&	-36.73 $ \pm$ 6.52	&	11.37	&	2.54	&	0.24	&	7.48	\\
52002$^3$&		7.89	$ \pm$ 0.24	&	2.43	$ \pm$ 0.20	&	-51.50 $ \pm$ 2.66	&	11.16	&	3.00	&	0.08	&	10.28	\\
63256$^2$	&	12.71	$ \pm$ 0.31	&	1.59	$ \pm$ 0.14	&	-37.33 $ \pm$ 2.99	&	13.09	&	2.77	&	0.17	&	12.64	\\
77695	&		13.25	$ \pm$ 0.77	&		-	&			-		 &	10.70	&	3.17	&	0.08	&	11.54	\\
81854	&		10.32	$ \pm$ 0.72	&		-	&		-		&		10.78	&	2.57	&	0.17	&	9.52	\\
87128	&		13.58	$ \pm$ 1.37	&		-	&		-		&		11.22	&	2.61	&	0.19	&	5.79	\\
114544	&	11.26	$ \pm$ 0.40	&		-		&		-		&	14.62	&	2.61	&	0.02	&	14.16	\\
\hline	
\hline
$^1$Magnya region       	&	14.67 	$ \pm$ 2.32	&	1.71	$ \pm$ 0.19  &	-39.45 $ \pm$ 13.63	& 11.73 $ \pm$ 1.33 &		&		&	\\
$^2$Eunomia region	&	10.93	$ \pm$ 1.30	&	1.81 $ \pm$ 0.17	&	-35.23 $ \pm$ 9.61	& 11.96 $ \pm$ 1.26 &		&		&	\\
$^3$Eos region		&	8.30 	$ \pm$ 1.50	&	2.07 $ \pm$ 0.30	&	-39.49 $ \pm$ 10.63	& 11.38 $ \pm$ 0.81 &		&		& \\
\hline
All MOVs	&	10.75  $ \pm$ 2.30 		&	1.82 $ \pm$	0.25 &	-39.52 $ \pm$ 9.49  &11.72    $ \pm$ 	1.12&		&		&	\\
\hline
Vesta family	&	11.46  $ \pm$ 2.65		&	1.48 $ \pm$	0.14 &	-23.82 $ \pm$ 3.59	& 11.60  $ \pm$ 1.06	&		&		&	\\
S-type asteroids	&	10.78  $ \pm$ 2.18		&  1.10	 $ \pm$ 0.05	 &	 -8.25 $ \pm$ 3.37 	&  10.79 $ \pm$ 0.46	&		&		&	\\	
\hline													
\hline	
\end{tabular}
~\\
\raggedright
\smallskip
\end{table*}

While a  steeper slopeA could be indicative of weathered surfaces (Fulvio et al. 2012, Fulvio et al. 2016), it is known that different taxonomic classes show  a different increase of the visible slope going to higher phase angles $\alpha$ (Barucci et al. 2017), and targets observed at $\alpha > 20^\circ$ can be affected by a moderate reddening. 
For this reason,  we report in Fig. 2a \emph{slopeA} vs phase angle obtained for MOV asteroids considered in this work, together with data reported in literature for V-type asteroids and computed in Ieva et al. (2016).
We found that there is a correlation between \emph{slopeA} and phase angle for basaltic main belt asteroids, and that MOV targets seem to fit coherently inside this scenario. For comparison, we computed the linear correlation for 205 S-type objects of the S3OS2 survey (Lazzaro et al. 2004): the reddening effect seems in this case less pronounced, and the residuals obtained applying this linear relation to our MOV sample are higher, suggesting a worse fit.
We also discovered that V-type NEAs show a different and milder linear trend than vestoids and MOVs (Fig 2b). This could be a consequence of the tidal perturbations induced by close encounters with terrestrial planets and already invoked to explain the ``fresh'' unweathered surface grains shown by V-type NEAs with respect to all other V-types (Fulvio et al. 2016). However, it should be reminded that there are only a handful of V-type NEAs with visible spectra, and further data for this class of V-types are required.

    \begin{figure}
   \includegraphics[angle=0,width=15cm]{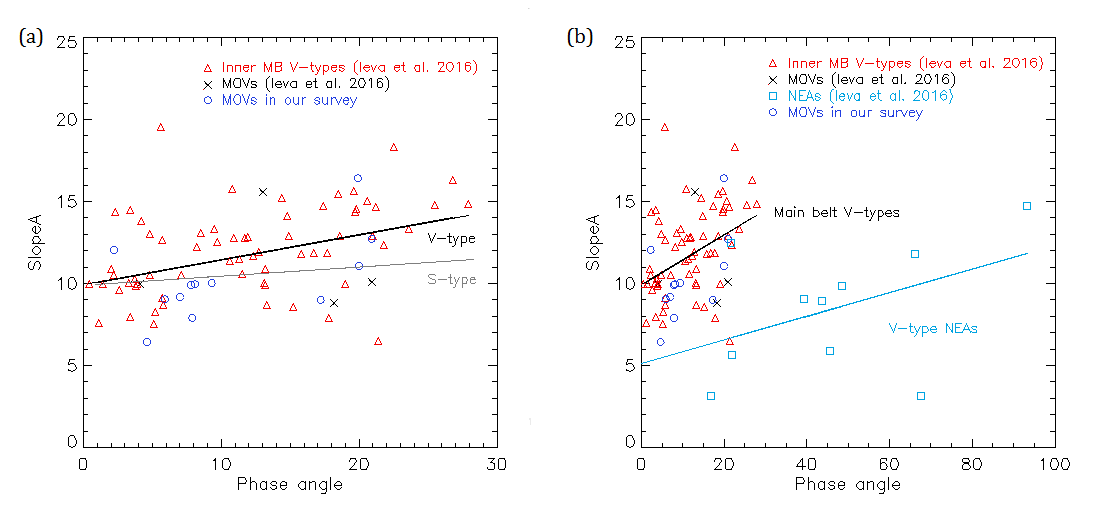}
      \caption{ a) The reflectivity gradient between 0.5-0.75  $\mu$m (\emph{slopeA}) as a function of the phase angle. The black line represent the linear fit obtained using all the basaltic asteroids considered here and in  Ieva et al. (2016), while the grey line represent the linear fit computed for 205 S-type asteroids reported in Lazzaro et al. (2004). It is possible to see that MOVs in our survey fit coherently among the V-type population. b) The same parameters compared with V-type NEAs. The latters show a less pronounced reddening due to the phase angle effect. This could be an indirect proof of their rejuvenated surfaces.  }
         \label{phaseangle}
   \end{figure}

Our recent statistical analysis (Ieva et al. 2016) has pointed out that the \emph{depth} computed for MOVs confirmed in literature is higher than objects belonging to the Vesta family, which can be interpreted as a different mineralogy (Cloutis et al. 2013).
In order to confirm if our final sample of 12 MOVs with computed \emph{depth} has a mineralogy compatible with Vesta and the Vesta family, first we used the relation obtained in Fig. 2a to correct  the computed \emph{slopeA} for phase angle effects.
Then, in Fig. 3a we reported \emph{slopeA} and \emph{depth} for MOV targets and the same parameters obtained for Vesta family objects, the ones that are physically and dynamically connected with Vesta. It is possible to see that vestoids and MOVs show a similar range of variation for  \emph{slopeA}, suggesting that they seem to have experienced similar amount of space weathering.
However, MOVs analyzed in our survey show a higher \emph{depth} than vestoids,  in agreement with MOVs previously analyzed, although few targets could be compatible with Vesta family, considering the large error bar in depth determination. 
For completeness, we also reported  in Fig. 3a the region occupied by the same spectral parameters computed for the sample of S-type objects from the S3OS2 survey (Lazzaro et al. 2004). 
The S-type population seem to occupy a different region in the diagram, showing an even lower \emph{depth}, which again is symptomatic for a different mineralogy.

       \begin{figure}
   \includegraphics[angle=0,width=16cm]{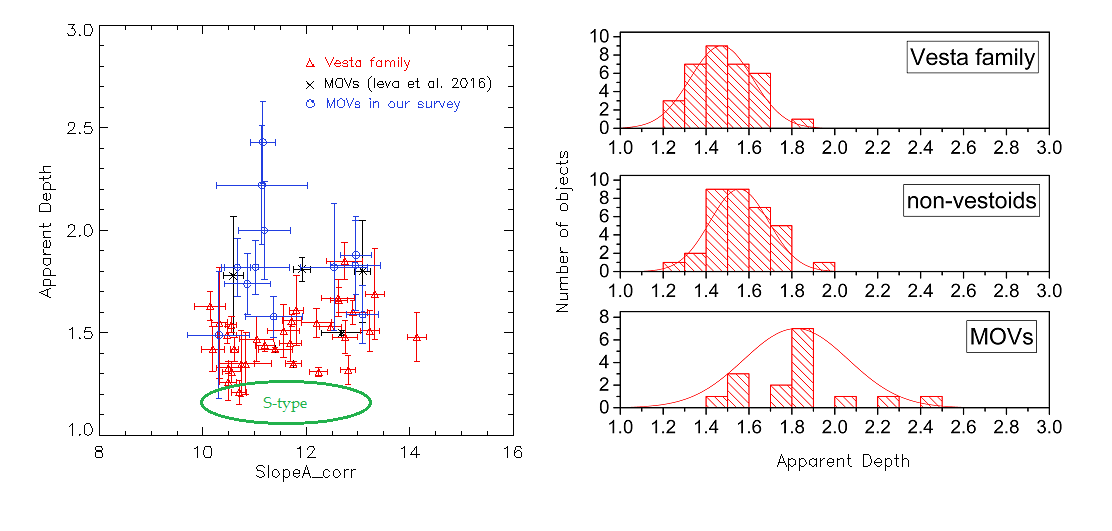}
      \caption{ a) The reflectivity gradient between 0.5-0.75  $\mu$m (\emph{slopeA}) vs the reflectance ratio at 0.75/0.9  $\mu$m (\emph{depth}) for asteroids belonging to the Vesta family and  MOVs reported in Ieva et al. (2016), together with spectral parameters computed for MOV objects in our sample using the same approach. We also reported in green the region occupied by the same spectral parameters computed for S-type asteroids. b) Histogram distribution of the \emph{depth} parameter for the three main populations considered in our statistical analysis: Vesta family, non-vestoids and MOVs. While the distributions for vestoids and non-vestoids seem to be centered at lower mean \emph{depth}, the MOV distribution is shifted towards higher \emph{depth}. The curves superimposed are the normal distributions with the same mean and standard deviation of the corresponding set of data.}
         \label{slope_depth}
   \end{figure}

The differences among Vesta family and MOVs are also shown in Fig. 3b, where we reported the histogram distributions for the \emph{depth} parameter of MOVs, vestoids and the whole population of non-vestoids analyzed in our previous statistical work: fugitives,  inner others (IOs) and low-i. For the description of these dynamical groups, we refer the reader to Ieva et al. (2016). Without making any theoretical assumption, we may say that the MOV distribution is more skewed towards higher values of \emph{depth} compared to the Vesta family and non-vestoids distribution.

In order to  quantify the difference between \emph{depth} values for the dynamical populations considered above we decided to apply a non-parametric statistical test. Values for Vesta family and non-vestoids were taken among V-type objects considered in Ieva et al. (2016) with available visible spectra  and computed \emph{depth}.
The Kolmogorov-Smirnov test seems more suitable due to the small size of the samples (33 objects for the Vesta family, 34 for the non-vestoid sample, 16 for MOVs). We used an online tool\footnote{http://www.physics.csbsju.edu/stats/} to compute the probability of the null hypothesis of no difference between the aforementioned datasets.
Results, reported in Tab. 3, show that the probability associated with this event is extremely small for both Vesta family and non-vestoids  (less than 0.1\%), meaning that we have a 99.9\% confidence level that the differences  we see between the MOV population and other V-types are real.
The same result happens if we confront the MOV sample with fugitives and low-i. No conclusion has been reached when comparing MOVs with IOs due to the limited sample of this dynamical group (only 6 objects).

\begin{table*}
       \caption{Kolmogorov-Smirnov test performed between the depth values computed for the MOV sample, and other dynamical groups of V-type asteroids in the main belt: Vesta family objects and fugitives, IOs and low-i, collectively called as non-vestoids. For an explanation of these groups, see Ieva et al. (2016)}
        \label{ks-test}
\begin{tabular}{|l|c|c|} \hline
\hline
MOVs vs&D	&	\emph{p}\\
\hline
Vesta family (33) & 0.7197 & <0.001 	\\
\hline
Fugitives (14) & 0.6786& 0.001	\\
Low i (14) & 0.6875 & 0.001		\\
\hline
Non-vestoids (34)& 0.6618 & <0.001 	\\
\hline													
\hline	
\end{tabular}

\textbf{NOTE:} \emph{p} is the probability, under the null hypothesis, of no difference between the two datasets. The smaller the \emph{p}, the more significant the test is. For 6 objects belonging to IOs no conclusion has been reached due to the limited size sample.\\
~\\
\raggedright
\smallskip
\end{table*}


\subsection{Orbital analysis}
Basaltic material in the main belt is extremely spread in the orbital plane. 
While the majority of V-type asteroids belong to the Vesta family, the presence of a cluster of basaltic material in a region dynamically inaccessible from Vesta (the middle and outer main belt) is a possible clue for the presence of other differentiated family(ies). The search of another basaltic progenitor has been inconclusive so far, although other dynamical families have been associated with basaltic material and differentiation, like Eunomia (Carruba et al. 2007, Nathues 2010) and Eos (Moth{\'e}-Diniz et al. 2008, Huaman et al. 2014).

We report in Fig. 4 the proper semi-major axis \emph{a} versus the proper inclination  \emph{i} for the Vesta, Eunomia and Eos family, MOVs analyzed in Ieva et al. (2016), MOVs featured in this article, and MOV candidates selected from various works (Masi et al. 2008,  Oszkiewicz et al. 2014).

       \begin{figure}
   \includegraphics[angle=0,width=15cm]{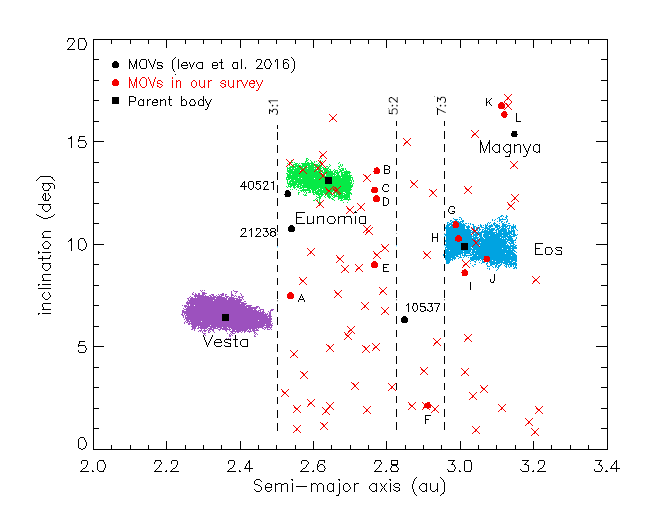}
      \caption{Proper semi-major axis vs proper inclination for objects belonging to Vesta (purple), Eunomia (green) and Eos family (cyan), together with MOV candidates (x) selected from several works (Masi et al. 2008, Oszkiewicz et al. 2014); MOVs with spectra available in literature and analysed in Ieva et al. (2016) are reported with a black filled dot, while MOVs spectra analysed in our work are indicated with a red filled dot. 
Legend: A) 48448, B) 23321, C) 63256, D) 22308, E) 24264, F) 47063, G) 41243, H) 52002, I) 14447, J) 10769, K) 27219, L) 14562}
         \label{orbital}
   \end{figure}

Three MOV targets presented in this work (48448, 24264 and 47063, respectively A, E and F in Fig. 4)  have orbital elements isolated from all the MOVs featured in our sample.
To look for a possible cluster of MOVs we identified three dynamical regions with a higher concentration of MOVs, obtained considering MOVs featured in this work and MOVs reported in Ieva et al. (2016). We found:

\begin{itemize}
\item 4 objects in the proximity of the Eunomia family (22308, 23321, 40521, 63256), with a semi-major axis $2.5 < a < 2.8$ and inclination $11^\circ  < i < 15^\circ$;  
\item 4 objects in the proximity of the Eos family (10769, 14447, 41243, 52002),  with semi-major axis $ a > 2.95$ and inclination $ i < 12^\circ$;
\item 2 objects in the proximity of Magnya (14562, 27219) and Magnya itself, with semi-major axis $ a > 2.95$ and inclination $ i > 14^\circ$.
\end{itemize}

Using the asteroid families computed on AstDyS\footnote{http://hamilton.dm.unipi.it/astdys/} (Milani et al. 2014)  with the Hierarchical Clustering Method (HCM), we identified that three new MOVs in the Eunomia region (23321, 63256 and 22308, respectively B, C and D in Fig. 4) do not belong to the Eunomia family, while three out of four newly basaltic objects identified in the proximity of Eos region (41243, 52002 and 10769, respectively G, H and J in Fig. 4) do belong to the Eos family. No asteroidal family seems to be associated with Magnya. However, among the two MOVs in its proximity, 27219 seems to belong to the Klumpkea/Tirela family, while 14562 seems to be a background object (K and L in Fig. 4).

We computed the average spectral parameters for MOVs belonging to those regions. Results, reported in Tab. 2, show that the three dynamical regions have similar \emph{slopeA} and \emph{slopeB}, with slight changes in the \emph{depth}. 
Finally, we stress that although new observations of MOVs in those dynamical ranges are desired  to enlarge the statistical sample, both \emph{depth} and \emph{slopeB} values computed for the MOV population are systematically higher than those obtained for the Vesta family.

\section{Discussion and Conclusions}
In our survey we have obtained new visibile spectra for 18 MOV candidates. All of them reside beyond 2.5 au, in a region dynamically inaccessible for objects excaped from the Vesta family.
We have retrieved spectral parameters using the same approach adopted in our previous statistical analysis. Our analysis has shown that the differences between MOVs and vestoids seem to be more evident comparing \emph{depth}, with the MOV population having the most extreme spectral parameters of the sample. MOVs show in particular a greater \emph{depth}  than the Vesta family and S-type asteroids.
While several authors (Burbine et al. 2001, Florczak et al. 2002) have already discussed that non-vestoids in the inner main belt seem to present a deeper 0.9 $\mu$m band, we point out that the average  \emph{depth} computed for the MOV population (1.82) indicate an even deeper band. Moreover, this value is still higher than those computed in Ieva et al. (2016) for the different non-vestoid populations: fugitives (1.62), IOs (1.53) and low-i (1.52). 
These differences are also clearly shown when we compared the histogram distributions for MOVs, vestoid and non-vestoid populations, with MOVs showing a distribution shifted towards higher  \emph{depth}.
Finally, we performed a statistical test to quantify the differences between those dynamical groups. Our results suggest, with a 99.9\% level of confidence, that the differences we see between MOVs and Vesta family (and also non-vestoids) are not due to chance.

The presence of a cluster of basaltic objects with similar spectral parameters, but different than Vesta's, could be a strong indicator of the presence of another differentiated family. 
The analysis of the spectral parameters in three different regions of the main belt with a high concentration of MOVs has shown that the \emph{depth} is always higher than the Vesta family. Moreover, three out of four MOVs in the proximity of Eos belong to the Eos family. 
Moth{\'e}-Diniz et al. (2008) mineralogically characterized members of the Eos family, concluding that they could be consistent with a partially differentiated body.
Burbine et al. (2017) suggested that the finding of basaltic asteroids in the Eos family could imply that Eos underwent a significant degree of differentiation. While the possibility for these objects to be interlopers of the family cannot be excluded, their presence is a further evidence toward the presence of another differentiated family in this region.
 
On the contrary, three MOVs in the proximity of Eunomia do not belong to its dynamical family. 
 This could be correlated to the respective age of the Eunomia and Eos family, with the first being older and more disperse in the orbital space (Spoto et al. 2015). Or these MOVs could be the remnants of another basaltic progenitor.

The presence of other differentiated families could alter our vision on the frequency of these processes in the early Solar System. Differentiation was most likely driven by the presence of $^{26}$Al and $^{60}$Fe, two radioactive isotopes. Bottke et al. (2006), based on the $^{26}$Al and $^{60}$Fe half-life period and their original location in the solar nebula, theorized that early differentiation could have been occurred only in the inner main belt ( a $<$ 2 au).
The proof of the existence of (an)other basaltic parent body beyond the 3:1 resonance could alter the current paradigma of differentiation processes, implying that the extension of the temperature gradient in the protosolar nebula at the epoch of planetary formation was different than always thought, in order to reach the right amount of heat able to sustain differentiation at solar distances \emph{a} $>$ 2.5 au. 
Another possibility is that basaltic progenitors  have formed inward and then scattered further away from the Sun.

Recent dynamical simulation (Brasil et al. 2017 and references therein) pointed out that, considering the effect of a ``jumping jupiter'' instability, it is indeed possible to implant basaltic material from Vesta to the middle and outer main belt in the early stages of Solar System. However, the implantation probability seem to be very low ($\sim 10\%$ in the middle belt and $\sim 1\%$ in the outer belt). Moreover, the authors don't seem to have an explanation for the current position of Magnya, whose origin still remain unknown.

With our survey we increase the number of basaltic objects identified in the middle and outer main belt by a factor 2. The sample of MOV candidates characterized is still limited to a few dozens, and the number of MOVs with a NIR spectroscopic characterization is even lower: only five.
New observations of MOV candidates, identified through the visible SDSS catalog or in the NIR range, like the VISTA catalogue (Licandro et al. 2017) are extremely needed, particularly in the NIR range,  in order to improve the current statistics on the MOV sample and better constrain the mineralogy. The incredible spectral resolution achieved by the DAWN mission on Vesta offer a suitable benchmark to compare its and MOVs NIR spectral parameters.
At the same time, new dynamical simulations are needed to confirm or reject the Vesta paternity of the MOV population.

\section*{Acknowledgements}
Support by CNPq (305409/2016-6) and FAPERJ (E-26/201.213/2014) is acknowledged by DL.
DF thanks the Brazilian foundation CNPq for financial support (``Bolsa de Produtividade em Pesquisa, PQ 2015'' - Processo: 309964/2015-6 - and ``Chamada Universal 2016'' - Processo: 426929/2016-0). 
DP has received funding from the European Union's Horizon 2020 research and innovation programme under the Marie Sklodowska-Curie grant agreement n. 664931.
We finally thank Ricardo Gil-Hutton for his useful and precious comments.





\section*{References}
Ammannito E., et al., 2013, Meteoritics and Planetary Science, 48, 2185 \\
Barucci M. A., et al., 2017, European Planetary Science Congress, 11, EPSC2017 \\
Binzel R. P., Xu S., 1993, Science, 260, 186 \\
Binzel R. P., Masi G., Foglia S., 2006, in AAS/Division for Planetary Sciences Meeting Abstracts \#38. p. 627 \\
Bland P. A., et al., 2009, Science, 325, 1525 \\
Bottke W. F., Nesvorn{\'y} D., Grimm R. E., Morbidelli A., O'Brien D. P., 2006, Nature, 439, 821\\
Brasil P. I. O., Roig F., Nesvorn{\'y} D., Carruba V., 2017, MNRAS, 468, 1236\\
Burbine T. H., Buchanan P. C., Binzel R. P., Bus S. J., Hiroi T., Hinrichs J. L., Meibom A., McCoy T. J., 2001, Meteoritics and Planetary Science, 36, 761\\
Burbine T. H., DeMeo F. E., Rivkin A. S., Reddy V., 2017, Evidence for Differentiation among Asteroid Families. pp 298{320, doi:10.1017/9781316339794.014 \\
Bus S. J., Binzel R. P., 2002, Icarus, 158, 146\\
Carruba V., Michtchenko T. A., Lazzaro D., 2007, A\&A, 473, 967\\
Carruba V., Huaman M. E., Domingos R. C., Santos C. R. D., Souami D., 2014, MNRAS, 439, 3168\\
Carvano J. M., Hasselmann P. H., Lazzaro D., Moth{\'e}-Diniz T., 2010, A\&A, 510, A43\\
Cloutis E. A., et al., 2013, Icarus, 223, 850\\
Cruikshank D. P., Tholen D. J., Bell J. F., Hartmann W. K., Brown R. H., 1991, Icarus, 89, 1\\
DeMeo F. E., Binzel R. P., Slivan S. M., Bus S. J., 2009, Icarus, 202, 160\\
Duffard R., Roig F., 2009, Planet. Space Sci., 57, 229\\
Florczak M., Lazzaro D., Duffard R., 2002, Icarus, 159, 178\\
Fulvio D., Brunetto R., Vernazza P., Strazzulla G., 2012, A\&A, 537, L11\\
Fulvio D., Perna D., Ieva S., Brunetto R., Kanuchova Z., Blanco C., Strazzulla G., Dotto E., 2016, MNRAS, 455, 584\\
Huaman M. E., Carruba V., Domingos R. C., 2014, MNRAS, 444, 2985\\
Ieva S., et al., 2014, A\&A, 569, A59\\
Ieva S., Dotto E., Lazzaro D., Perna D., Fulvio D., Fulchignoni M., 2016, MNRAS, 455, 2871\\
Lazzaro D., 2009, in Revista Mexicana de Astronomia y Astrosica Conference Series. pp 1-6\\
Lazzaro D., et al., 1999, Icarus, 142, 445\\
Lazzaro D., et al., 2000, Science, 288, 2033\\
Lazzaro D., Angeli C. A., Carvano J. M., Moth{\'e}-Diniz T., Duffard R., Florczak M., 2004, Icarus, 172, 179\\
Licandro J., Popescu M., Morate D., de L{\'e}on J., 2017, A\&A, 600, A126\\
Marchi S., et al., 2012, Science, 336, 690\\
Masi G., Foglia S., Binzel R. P., 2008, in Asteroids, Comets, Meteors 2008. p. 8065\\
McCord T. B., Adams J. B., Johnson T. V., 1970, Science, 168, 1445\\
Milani A., Cellino A., Knezevic Z., Novakovic B., Spoto F., Paolicchi P., 2014, Icarus, 239, 46\\
Moth{\'e}-Diniz T., Carvano J. M., Bus S. J., Duffard R., Burbine T. H., 2008, Icarus, 195, 277\\
Nathues A., 2010, Icarus, 208, 252\\
Oszkiewicz D. A., Kwiatkowski T., Tomov T., Birlan M., Geier S., Penttila A., Poliinska M., 2014, A\&A, 572, A29\\
Roig F., Gil-Hutton R., 2006, Icarus, 183, 411\\
Roig F., Nesvorn{\'y} D., Gil-Hutton R., Lazzaro D., 2008, Icarus, 194, 125\\
Scott E. R. D., Greenwood R. C., Franchi I. A., Sanders I. S., 2009, Geochimica Cosmochimica Acta, 73, 5835\\
Spoto F., Milani A., Knezevic Z., 2015, Icarus, 257, 275\\
Tholen D. J., Barucci M. A., 1989, in Binzel R. P., Gehrels T., Matthews M. S., eds, Asteroids II. pp 298-315\\
Thomas P. C., Binzel R. P., Gaffey M. J., Storrs A. D., Wells E. N., Zellner B. H., 1997, Science, 277, 1492\\





%
%


\bsp	
\label{lastpage}
\end{document}